\documentclass[lettersize,journal]{IEEEtran}
\usepackage{amsmath,amsfonts}
\usepackage{algorithmic}
\usepackage{algorithm}
\usepackage{array}
\usepackage[caption=false,font=normalsize,labelfont=sf,textfont=sf]{subfig}
\usepackage{textcomp}
\usepackage{stfloats}
\usepackage{url}
\usepackage{verbatim}
\usepackage{graphicx}
\usepackage{cite}
\usepackage{cancel}
\usepackage{xcolor}
\renewcommand{\baselinestretch}{0.997}

\begin{document}

\title{A Joint Reconstruction-Triplet Loss Autoencoder Approach Towards Unseen Attack Detection in IoV Networks\thanks{This material is based upon work supported by the National Science Foundation under Grant Numbers CNS-2318726, and CNS-2232048, and Clemson University R-initiative funding.
This work was supported by Clemson University’s Virtual Prototyping of Autonomy Enabled Ground Systems (VIPR-GS), under Cooperative Agreement W56HZV-21-2-0001 with the US Army DEVCOM Ground Vehicle Systems Center (GVSC).
DISTRIBUTION STATEMENT A. Approved for public release; distribution is unlimited. OPSEC\#9299
}}

\author{Julia Boone,~\IEEEmembership{Graduate Student Member,~IEEE}, Tolunay Seyfi, Fatemeh Afghah,~\IEEEmembership{Senior Member,~IEEE}
        % <-this % stops a space
\thanks{J. Boone, T. Seyfi, and F. Afghah are with the  Holcombe Department of Electrical and Computer Engineering at Clemson University. Emails: \{jcboone, tseyfi, fafghah\}@clemson.edu.

% Copyright (c) 2025 IEEE. Personal use of this material is permitted. However, permission to use this material for any other purposes must be obtained from the IEEE by sending a request to pubs-permissions@ieee.org.
}% <-this % stops a space
\thanks{Manuscript received September 15, 2024; revised May 5, 2025; accepted May 21, 2025.}
}

\markboth{IEEE INTERNET OF THINGS JOURNAL,~Vol.~X, No.~X, XXXX~XXXX}%
{Shell \MakeLowercase{\textit{et al.}}: A Sample Article Using IEEEtran.cls for IEEE Journals}

\IEEEpubid{
% 0000--0000/00\$00.00~\copyright~2021 IEEE
}

\maketitle

\begin{abstract}
Internet of Vehicles (IoV) systems, while offering significant advancements in transportation efficiency and safety, introduce substantial security vulnerabilities due to their highly interconnected nature. These dynamic systems produce massive amounts of data between vehicles, infrastructure, and cloud services and present a highly distributed framework with a wide attack surface. In considering network-centered attacks on IoV systems, attacks such as Denial-of-Service (DoS) can prohibit the communication of essential physical traffic safety information between system elements, illustrating that the security concerns for these systems go beyond the traditional confidentiality, integrity, and availability concerns of enterprise systems. Given the complexity and volume of data generated by IoV systems, traditional security mechanisms are often inadequate for accurately detecting sophisticated and evolving cyberattacks. Here, we present an unsupervised autoencoder method trained entirely on benign network data for the purpose of unseen attack detection in IoV networks. We leverage a weighted combination of reconstruction and triplet margin loss to guide the autoencoder training and develop a diverse representation of the benign training set. We conduct extensive experiments on recent network intrusion datasets from two different application domains, industrial IoT and home IoT, that represent the modern IoV task. We show that our method performs robustly for all unseen attack types, with roughly 99\% accuracy on benign data and between 97\% and 100\% performance on anomaly data. We extend these results to show that our model is adaptable through the use of transfer learning, achieving similarly high results while leveraging domain features from one domain to another.
\end{abstract}


\section{Introduction}

\IEEEPARstart{V}{ehicular} networks help realize real-time communications between vehicles, infrastructure, and other essential components of a transportation system. The onset of Internet of Things (IoT) systems marks the implementation of data-rich, persistent interconnections between various sensing devices designed for the monitoring and automation of various tasks. From industrial monitoring mechanisms designed to detect production deficiencies \cite{iiot} to battlefield warnings generated via collected data \cite{mil_iot}, IoT systems provide key ways in which safety and efficiency can be increased within a variety of systems. The Internet of Vehicles (IoV) has emerged from the IoT as a way by which we can create intelligent transportation systems (ITS) to provide this persistent and data-rich inter-connectivity between vehicles. Despite the advantages of such systems, security for the IoV is an open challenge \cite{iov_security}. Given the physical safety and sensitive data risks that can be caused by attacks on interconnected vehicles, it is paramount that these systems are secured using methods capable of robustly detecting attacks.

Attacks on distributed and decentralized systems like interconnected vehicles can have widescale impacts. In late 2016, a botnet called Mirai designed for Distributed Denial of Service (DDoS) attacks was utilized to take down a variety of websites or online services, such as the security journalist Brian Krebs and the Dynamic DNS provider Dyn \cite{mirai,mirari2}. Mirai was able to achieve this by targeting IoT devices with ARC processors running the Linux operating system and attempting to log into the device with default credentials. An estimated half a million IoT devices were utilized to carry out Mirai's attacks and, in the case of Krebs' attack, an estimated \$323,973 of costs were inflicted on device owners in considering energy and bandwidth costs \cite{miraicost}.

Generalizability is a particular concern in attack detection systems. While some methods are well suited for particular attacks or environments, a robust anomaly detection method should be capable of detecting various attacks in different environments with high degrees of accuracy. One key issue underneath this umbrella of generalizability in practical applications of anomaly detection methods is unseen attack detection to mitigate zero-day, or unknown, attacks. Ideally, an anomaly detection method should be able to detect new attack types as deviations from the norm when they occur. The Mirai attack is a key example of this need as it was an unknown attack capable of operating undetected on a massive number of decentralized devices. The ability for a model to not only perform well on an individual dataset but to perform well across datasets and to be generalizable through domain adaptation methods, such as transfer learning, is critical in the application of attack detection models across different environments.

While machine learning (ML) methods perform well on labeled sets of network data, they typically fail to detect data unseen during their training process and struggle to specifically capture the fine-grain spatial and temporal features of the input data \cite{machines_dlvsml,fawaz2019deep} From this, various artificial intelligence (AI)-based methods have been developed for network attack detection. Some pre-existing intelligent approaches to the anomaly detection task are entirely supervised, where all of the incoming data stream is labeled as anomalous or benign. While supervised methods generally achieve high performance, they are unrealistic for the network security task as raw traffic flows are not inherently labeled as benign or malicious that may not have pre-existing data for such new attacks. It can be time-consuming, expensive, and potentially infeasible to collect real attack data in new domains, such as an IoV scenario where vehicles operate in a highly dynamic environment that cannot be fully modeled ahead of time, creating a complicated and unpredictable attack landscape. In this scenario, it is clear that being able to leverage known benign data and/or the performance of a model from another domain in a new domain is critical in ensuring the safety of newly deployed systems with no pre-existing attack knowledge.

To this end, this work focuses on the development of unsupervised anomaly detection towards unseen attack detection. We develop an unsupervised autoencoder-based approach, where the model is trained on the aggregated sets of benign network flows only and reconstruction error is leveraged as the anomaly metric. In using reconstruction error alone, however, the latent space representations of the benign class may be too intertwined with the anomalies, leading to the reconstruction of anomalous samples as benign. Additionally, reconstruction error alone may struggle to capture discriminative features necessary for the detection task as it is intended to capture general patterns. Inspired by existing contrastive learning approaches to the anomaly detection task, we modify the traditional reconstruction-based autoencoder training to address this by including a triplet margin loss to strengthen the latent space representation of the benign set from the encoder. This loss allows us to address the previous issues by introducing additional samples that represent similarity and dissimilarity into the model loss to both capture more relevant features of the benign class, which in turn helps strengthen the latent space representation boundaries of the benign class. We can also extend this diversification of the latent space through this loss as it relates to domain transference, in which clearly defined boundaries between the anomaly and benign classes out of a particular domain's training can then be leveraged in a target domain. It allows us to easily adapt the model to that new domain without needing to conduct the full training process, as we show with the use of transfer learning in our evaluations.

To the best of our knowledge, this work is the first in the IoV domain to perform entirely unsupervised anomaly detection via only training on the benign sample and utilizing two datasets that are application distinct from one another while also being well representative of modern IoV traffic patterns. This is especially critical in the context of zero-day attack detection, as every attack is effectively treated as a zero-day attack by our method, given that no pre-existing attack data is used for training. We show that  our model is high performing despite the application differences on these datasets and is not specialized to detecting any one attack type, instead being capable of detecting a wide breadth of potential attacks. Considering that the transference of attack knowledge from domain-to-domain is also highly important for zero-day attack detection in new environments, we also present results showing that our model is adaptable across application domains through the use of transfer learning.

Our contributions can be summarized as follows:
\begin{itemize}

    \item We develop a joint reconstruction error-triplet loss  based approach for an autoencoder for attack detection in IoT networks. This method trains the autoencoder to reconstruct the feature sequences of network flow data with a high degree of accuracy without overfitting to the benign set via the addition of the triplet loss. 
    \item We present a novel method specifically for the task of unseen attack detection in IoV networks. By training entirely on the benign set of traffic data, our method is entirely unsupervised and performs detection independent of the specific attack occurring, instead focusing on capturing the behaviors of the benign system traffic and detecting deviations. As such, our method is highly suitable for the unseen attack detection task in IoV networks.
    \item We evaluate our method on two recent and distinct datasets that well capture the traffic patterns of IoT networks. We argue that these datasets are better representatives of modern IoV problems compared to commonly used intrusion datasets across the current literature and have been underutilized in the development of robust network intrusion detection mechanisms in IoV works.
    \item We also illustrate the capabilities of our model as a generalizable method towards unseen attack detection in \textit{new} IoV environments via transfer learning. This work is one of the first to explore this capability of an anomaly detection specific model while using data that is appropriate for the modern IoV task.
\end{itemize}

\section{Related Works}

% Tolunay said he'd update this on Friday so will leave for him for now

\subsection{ML versus AI for AD}In the context of IoT systems, we specifically consider the task of time-series anomaly detection, which is critical for maintaining the security and functionality of these systems. Machine learning (ML) methods are commonly used for this task \cite{wustl-2021-paper,dromard2016online,hoang2019detection}. While ML performs well for some datasets, as data grows larger and more complex, ML may fail to catch the unique patterns within the data. Because deep learning (DL) can likely capture such patterns to outperform ML approaches \cite{pang2021deep}, DL approaches have become popular for the development of new anomaly detection methods. 

Unsupervised DL methods are broadly classified into two categories: prediction-based and reconstruction-based mechanisms. In prediction-based methods, regression models are trained on historical data to forecast future values of the system \cite{cook2019anomaly}. If the observed values deviate significantly from the predictions, they are considered to be anomalous. \cite{wu2019lstm} proposes a joint LSTM-Gaussian Naive Bayes model for industrial IoT (IIoT) anomaly detection, leveraging LSTM's forecasting capabilities and Gaussian Naive Bayes for outlier detection. This work leverages the forecasting capabilities of the LSTM model in order to generate the future time predictions and the Gaussian Naive Bayes model to perform outlier detection on the prediction error. Similarly, {\cite{saurav2018online}} uses a GRU-based RNN for online anomaly detection, accounting for natural shifts in the data distribution.

In contrast, reconstruction-based methods involve training generative models, such as autoencoders or GANs, on benign data to learn the normal data distribution. These models, once trained, use the learned distribution to reconstruct new data samples. Any significant deviation in the reconstruction error indicates an anomaly. \cite{provotar2019unsupervisedad} discusses the use of LSTM-based autoencoders for this purpose, while \cite{li2023anomaly} explores the application of GANs. Anomaly thresholds can be set either as fixed numerical values or based on dynamic statistical measures of the loss distribution, as demonstrated by \cite{jia2022dynamic}.

\subsection{Contrastive Learning for AI-Based AD and Its Relevance to IoV Networks} Building on these techniques, contrastive learning has emerged as a self-supervised approach that aims to extract meaningful representations from unlabeled data using proxy tasks. This method has gained attention for its ability to learn transformation-invariant representations, making it highly effective for unsupervised representation learning. By contrasting different views of the same sample (positive pairs) against views from different samples (negative pairs), contrastive learning enhances the model’s capacity to distinguish and understand data patterns. \cite{adversarial_contrastive} proposes an Adversarial Contrastive Autoencoder to improve multivariate time series anomaly detection by learning transformation-invariant representations through adversarial training. Positive and negative sample pairs are generated using multi-scale timestamp masks and random sampling. A 1D-CNN-based encoder extracts latent features from the samples, and composite features are created from positive and negative sample pairs while a discriminator decomposes these features. \cite{contrastive_autoencoder} focuses on multi-grained contrasting and data augmentation by integrating contrastive learning into an autoencoder framework for anomaly detection, leveraging both window-level and pixel-level contrastive tasks to learn normal patterns. An LSTM decoder is utilized for data reconstruction and calculation of anomaly scores based on reconstruction errors. Contextual and instance contrasting are combined with attention mechanisms to learn temporal features and invariant features from augmented views.

In IoV, however, reconstruction-based anomaly detection has been widely applied, with autoencoders leveraging reconstruction loss to identify sensor anomalies \cite{deepCAE2022}, detect anomalous driving behaviors \cite{saberVAE2023}, and pinpoint location inconsistencies in CAVs \cite{cavAnomaly2019}. These methods rely solely on reconstruction quality without enforcing clear separation between normal and anomalous samples. Contrastive loss is well-suited for this dynamic environment as it learns representations based on similarity relationships rather than fixed decision boundaries. By continuously adapting to shifting network conditions and mobility patterns, it ensures that normal behaviors remain well-clustered while anomalies, even subtle or context-dependent ones, are effectively separated. This flexibility makes contrastive learning particularly robust against the inherent variability of IoV networks. 

\subsection{Adaptive Anomaly Detection in Vehicle Systems}
Vehicular networks, part of the emerging IoV, face numerous challenges related to security and anomaly detection. Due to the real-time data exchange between vehicles and infrastructure, anomaly detection mechanisms must account for domain shifts across different environments, vehicle types, and driving conditions. This is particularly important in identifying malicious activities or system failures that may affect vehicle performance or safety. Several significant works have addressed these challenges (Table~\ref{tab:relatedworkstable}), and we will examine them in relation to our contribution. By putting these existing solutions into context with our own, we aim to show how our approach advances the field by enhancing domain adaptation and security in IoV systems.

{
\begin{table*}[ht]
\centering
\caption{Comparison of Relevant Works in Anomaly Detection for VANETs and Autonomous Driving Systems }
\label{tab:relatedworkstable}
\resizebox{\textwidth}{!}{%
\begin{tabular}{|l|l|l|l|l|}
\hline
\textbf{References} & \textbf{Learning Method} & \textbf{Dataset} & \begin{tabular}[l]{@{}c@{}}\textbf{ Supports Domain} \\ \textbf{Shift?}\end{tabular} & \textbf{Attack Types} \\ \hline
Nazat et al.\cite{XAI} & Supervised  & VeReMi, Sensor Dataset & No & DoS, Sybil Attacks, Message Falsification \\ \hline
ALMahadin et al. \cite{VANET_GRU} & Semi-supervised & NSL-KDD & Yes & DDoS, Phishing Attacks, Password Attacks, R2L, U2R\\ \hline
Nissar et al. \cite{VANET_VAE} & Unsupervised & NSL-KDD & Yes &  DDoS, Phishing Attacks, Password Attacks, R2L, U2R  \\ \hline
This paper & Unsupervised& ACI-2023, WUSTL-2021 & Yes & DoS, SQL Injection, Reconnaissance, Backdoor, Dictionary Brute Force, ARP Spoofing\\ \hline
\end{tabular}%
}
\end{table*}
}

%\ts{Should I add another paragraph highlighting our work in this subsection or is this version fine? I am concerned it will extend our page limit since we already do without it.}

% \begin{table*}[ht]
%     \centering
%     \caption{Comparison of Relevant Works in Anomaly Detection for VANETs and Autonomous Driving Systems \fa{i) our paper should be add as the last row. ii) the format of table does not seem right.  }}
%     \begin{tabular}{|p{1.5cm}|p{1.5cm}|p{1cm}|p{1cm}|p{1.5cm}|}
%         \hline
%         \textbf{Author(s)} & \textbf{Supervised/ Unsupervised} & \textbf{Dataset Type} & \textbf{Supports Domain Shift?} & \textbf{Number of Attacks}\\ 
%         \hline
%         \textbf{Nazat et al.\cite{XAI}} & Supervised & VeReMi, Sensor Dataset & No & DoS, Sybil Attacks, Message Falsification\\
%         \hline
%         \textbf{ALMahadin et al. \cite{VANET_GRU}} & Semi-supervised & NSL-KDD & Yes & DDoS, Phishing Attacks, Password Attacks, R2L, U2R \\ 
%         \hline
%         \textbf{Nissar et al. \cite{VANET_VAE}} & Unsupervised & NSL-KDD & Yes & DDoS, Sybil Attack, Malware, Privacy Attacks, Integrity, Availability Attacks\\
%         \hline
%     \end{tabular}\hspace{1cm}
%     \label{tab:anomaly-table}
% \end{table*}

One approach, as discussed in \cite{XAI}, focuses on detecting anomalies in Vehicular Ad-hoc Networks (VANETs) using supervised AI models while addressing the challenge of their "black-box" nature. To enhance the transparency of these models, the framework integrates two key explainability techniques: Shapley Additive Explanations (SHAP) and Local Interpretable Model-agnostic Explanations (LIME). SHAP provides global insights by quantifying the contribution of each feature to the overall model predictions, while LIME offers local interpretations by explaining individual model decisions on a per-sample basis.

%This dual approach improves both the transparency and trustworthiness of the system, allowing users to understand not only the general behavior of the model but also the rationale behind specific decisions, which is crucial in safety-critical vehicular environments.

The framework is evaluated on two real-world autonomous driving datasets: the VeReMi dataset and a custom sensor dataset. The VeReMi dataset, specifically designed for misbehavior detection in VANETs, simulates various cyber-attacks, including DoS, Sybil attacks, and message falsification, covering a total of 225 scenarios. Complementing VeReMi, the sensor dataset captures vehicular behaviors using data from ten distinct sensors, recording parameters such as location, speed, lane alignment, and headway time. These diverse features enable the model to detect abnormal vehicle behaviors that may indicate cyber-attacks or system malfunctions.
In the context of anomaly detection, the framework is tested against five different types of attacks, encompassing both traditional attack types (like DoS) and more specific vehicular misbehavior attacks.

%The integration of SHAP and LIME plays a crucial role in identifying and ranking the most significant features contributing to the detection of these anomalies. For instance, in the VeReMi dataset, vehicle position and speed along the x and y axes are highlighted as key indicators of abnormal behavior, while in the sensor dataset, lane alignment and consistency in sensor readings prove essential for detecting irregularities.
In addition to this framework, \cite{VANET_GRU} proposes an anomaly detection model for VANETs using a GRU-based deep learning architecture. The model introduces a semi-supervised technique called SEMI-GRU, which integrates GRU neural networks with the Synthetic Minority Oversampling Technique (SMOTE) oversampling technique to improve anomaly detection accuracy and reduce false positives. The GRU architecture is particularly advantageous for capturing long-term dependencies in sequential data while using fewer parameters than traditional Long Short-Term Memory (LSTM) networks, resulting in faster training. Furthermore, combining GRU with feed-forward neural networks (FNN) enhances feature extraction, leading to more refined anomaly detection.
The SEMI-GRU method addresses key challenges in anomaly detection, such as handling imbalanced datasets and detecting unknown cyber-attacks in VANET traffic. To combat class imbalance, the model employs the SMOTE, which generates synthetic samples for underrepresented attack types in the dataset. %This approach is especially effective for improving detection accuracy in rare categories, such as Remote-to-Local (R2L) and User-to-Root (U2R) attacks. 
The model is evaluated using the NSL-KDD dataset [24], which contains 42 features and several types of network attacks, including Denial of Service (DoS), Probe, Remote-to-Local (R2L), and User-to-Root (U2R). 
%By leveraging SMOTE, the SEMI-GRU model significantly enhances detection performance, particularly for R2L and U2R attacks, where oversampling creates a more balanced training set, leading to more reliable anomaly detection in VANET traffic.
Furthermore, \cite{VANET_VAE} presents an unsupervised approach for anomaly detection in VANETs using Variational Autoencoders (VAEs) optimized with multi-objective evolutionary algorithms, such as AGE-MOEA and R-NSGA-III. This framework focuses on detecting zero-day attacks and handling high-dimensional vehicular network traffic, making it particularly suitable for dynamic and evolving VANET environments. By learning latent data representations, the VAE model is capable of identifying novel intrusions without needing labeled data, which addresses one of the key limitations of supervised models.
Unlike \cite{VANET_GRU} that used the NSL-KDD dataset primarily to handle class imbalances in a semi-supervised setup, this framework employs the same dataset but focuses on unsupervised anomaly detection, leveraging the entire feature set to optimize detection accuracy across multiple objectives.
\vspace{4pt}

\section{IoV Network Traffic Datasets}

Analyzing the landscape of intrusion detection in the Internet of Vehicles (IoV), we observe a significant shortage of publicly available datasets explicitly designed for vehicular network security. Many works resort to using general IoT datasets as proxies for IoV traffic, which can be problematic given the fundamental differences between traditional IoT and vehicular environments. IoV systems introduce unique temporal patterns, mobility constraints, and attack surfaces, making it crucial to carefully assess dataset applicability.

Unlike traditional network intrusion datasets, such as NSL-KDD and its predecessor KDD Cup'99, IoV datasets must capture both network-based and mobility-induced anomalies. As depicted in Table \ref{tab:dataset_descriptions}, NSL-KDD refines the KDD Cup’99 dataset by reducing redundancy and mitigating class imbalance issues. It has removed redundant and duplicate records to avoid biased learning and overfitting during model training, leading to faster and computationally more feasible model training and evaluation while still retaining the complexity needed for network intrusion detection tasks. Unlike the KDD Cup’99 dataset, it does not suffer from class imbalance, as certain attack types were overrepresented, which hindered the ability of models to generalize well, particularly for underrepresented attack types. Despite these improvements, NSL-KDD and KDD Cup’99 do not entirely capture more modern attack types or the dynamic nature of traffic found in current IoV systems. Additionally, while NSL-KDD addressed class imbalance to some extent, challenges remain with underrepresented attack categories such as Remote-to-Local (R2L) and User-to-Root (U2R) attacks.

To address these limitations, datasets such as VeReMi have been developed specifically for vehicular anomaly detection. Unlike NSL-KDD, VeReMi integrates spatiotemporal features, such as GPS data, vehicle speed, and direction, making it highly relevant for detecting falsified positioning data and misbehavior attacks in vehicular networks. The VeReMi dataset, used in more recent vehicular network research (e.g., \cite{XAI}), focuses on threats unique to IoV, including Denial-of-Service (DoS) attacks on vehicular messaging, Sybil attacks where multiple fake vehicles compromise decision-making, and message falsification attacks that could manipulate vehicle responses. The dataset encompasses 225 distinct attack scenarios, providing a richer representation of adversarial behaviors in vehicular networks. However, VeReMi primarily focuses on vehicle-to-vehicle (V2V) communication anomalies and does not adequately capture network-side attacks, where adversaries exploit vulnerabilities within V2X infrastructure (e.g., roadside units (RSUs), edge servers, or core network elements) to launch large-scale disruptions.

For addressing network-side attacks, we find that the ACI-IoT-2023 and WUSTL-IIoT-2021 datasets provide broader coverage of IoV-relevant cyber threats compared to VeReMi and NSL-KDD. ACI-IoT-2023 includes brute-force and Address Resolution Protocol (ARP) spoofing attacks, which target authentication and network integrity—key components of V2X security. ARP spoofing, for instance, can redirect vehicle communication within IoV networks, potentially leading to man-in-the-middle attacks where adversaries can modify or intercept safety-critical messages. Similarly, WUSTL-IIoT-2021 introduces command injection and reconnaissance attacks, both of which are highly relevant to securing real-time vehicle control systems against unauthorized commands and stealthy data gathering.

A key limitation of existing vehicular anomaly detection datasets is that they often assume an attacker operates externally, either by injecting fake GPS signals or spoofing nearby vehicles. However, a sophisticated attacker could compromise the IoV network itself, blending into the system while executing malicious actions at strategically critical moments. For example, an adversary could send falsified GPS data to the V2X infrastructure, misleading the network into believing a vehicle is in a different location. At a later time, they could remotely manipulate vehicle commands, such as triggering unintended acceleration or disabling braking systems, leading to catastrophic safety failures.

To fully model network-based IoV attacks, datasets must encompass not only message anomalies but also network-layer attacks where adversaries leverage V2X communication channels to systematically manipulate vehicular behavior. While ACI-IoT-2023 and WUSTL-IIoT-2021 capture network-oriented attack types, further work is needed to bridge the gap between mobility-driven threats and network-layer intrusions. The next generation of IoV anomaly detection datasets should integrate real-time network telemetry, vehicular control data, and multimodal sensor fusion to improve the detection of stealthy, coordinated attacks within V2X ecosystems.

By structuring the discussion around these key differences, we highlight why IoV anomaly detection presents challenges distinct from traditional time-series datasets. The combination of dynamic mobility, real-time constraints, cross-layer attacks, and adversarial deception techniques makes IoV security an evolving research challenge that demands novel detection mechanisms beyond conventional network intrusion models.

\begin{figure}[ht!]
    \centering
    \includegraphics[width=0.70\linewidth]{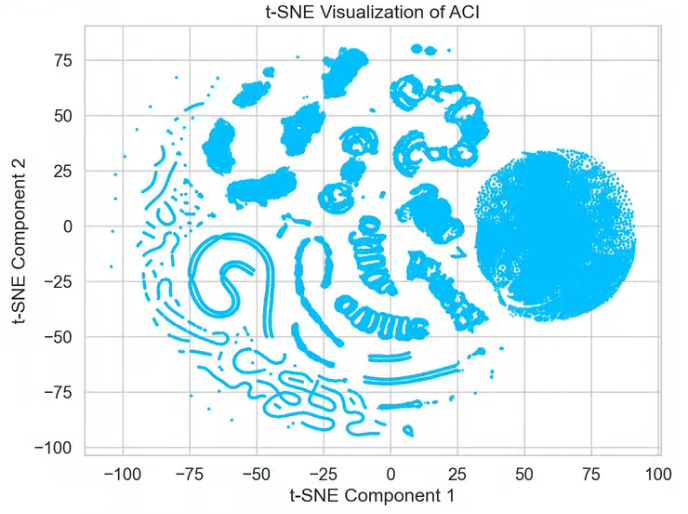}
    \makeatletter
    \renewcommand{\@makecaption}[2]{\centering \small #1. #2\par}
    \makeatother
    \caption{t-SNE visualization of the ACI-IoT-2023 dataset}
    \label{fig:tsne_aci}
\end{figure}

Figure \ref{fig:tsne_aci} presents the t-SNE visualization of the ACI-IoT-2023 dataset's network flows, revealing a highly clustered and heterogeneous structure, with distinct regions of data points interspersed with well-separated groups. This suggests a dataset containing diverse traffic patterns, likely representing a wide range of attack behaviors and benign communications. The clear separation of clusters indicates that network activities exhibit distinct behavioral patterns, which aligns with ACI-IoT-2023's inclusion of authentication-based threats like brute force attacks and ARP spoofing. These types of attacks can create sharp deviations in feature space, reinforcing the importance of anomaly detection models that can effectively distinguish between normal and compromised network states.

\begin{figure}[ht!]
    \centering
    \includegraphics[width=0.70\linewidth]{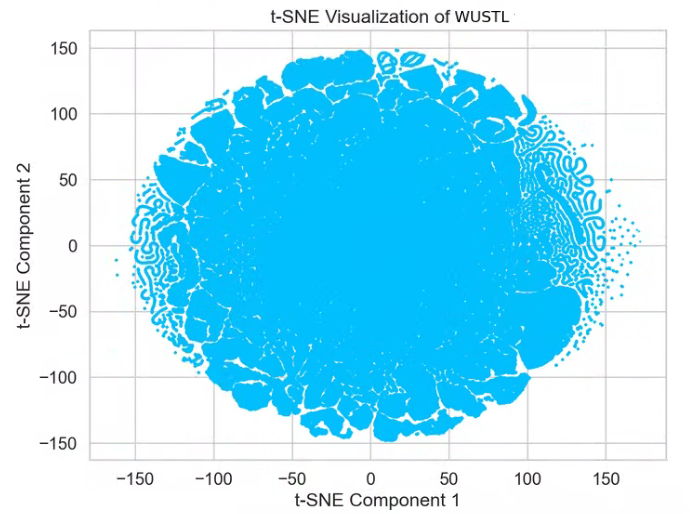}
    \makeatletter
    \renewcommand{\@makecaption}[2]{\centering \small #1. #2\par}
    \makeatother
    \caption{t-SNE visualization of the WUSTL-2021 dataset}
    \label{fig:tsne_wustl}
\end{figure}

Conversely, Figure \ref{fig:tsne_wustl} depicts the t-SNE visualization of the WUSTL-IIoT-2021 dataset's network flows, which displays a more continuous and densely packed distribution of data points. Unlike ACI-IoT-2023, this dataset appears to have less distinct clustering, suggesting that the data contains gradual variations between normal and anomalous traffic patterns, rather than sharply delineated attack signatures. This structure aligns with command injection and reconnaissance attacks, which can be more subtle and progressively influence network behavior over time. The overlapping nature of the data also suggests that anomalies in this dataset may be more challenging to detect, requiring methods capable of learning fine-grained distinctions within benign and malicious activity.

The differences between ACI-IoT-2023 and WUSTL-IIoT-2021 emphasize the need for robust anomaly detection frameworks that can handle both highly clustered, distinct attack patterns (ACI-IoT-2023) and continuous, stealthy attack behaviors (WUSTL-IIoT-2021). By leveraging advanced contrastive learning and domain adaptation techniques, we can develop intrusion detection methods that generalize effectively across these varying IoV environments.

\begin{table*}[h!]
    \centering
    \caption{IoT Intrusion Datasets}
    \begin{tabular}{|c|c|c|c|c|c|}
        \hline
         \textbf{Dataset} & \textbf{Observation \#} & \textbf{Attacks} & \textbf{Duration} & \textbf{Device \#} & \textbf{Year}\\
         \hline
         KDD Cup 99 & 4,418,358 & DoS, Probe, R2L, U2R & 9 weeks & unknown & 1999\\
         \hline
         NSL-KDD & 160,367 & DoS, Probe, R2L, U2R & 9 weeks (subset of KDD Cup 99) & unknown & 2009\\
         \hline
         WUSTL-2021 & 1,194,464 & SQL Injection,
         DoS, Recon, Backdoor & 53 hours & 10 & 2021\\
         \hline
         ACI-IoT-2023 & 1,231,411 & DoS, Recon, Brute Force & 5 Days & 49 & 2023\\
         \hline
    \end{tabular}
    \label{tab:dataset_descriptions}
\end{table*}

% \jb{do we need to include a table for NSL-KDD vs KDD cup explicitly? I can add them both to the table in evaluation that has WUSTL and ACI right now or move that table up here, but maybe can save space if we put them all in one table} \fa{Yes, one table (spanning two columns of paper) for all datasets in the new dataset section would be better. }

% \begin{table}[ht]
% \centering
% \caption{Comparison of Total Instances between NSL-KDD and KDD Cup '99 Datasets}
% \begin{tabular}{|p{2.4cm}|p{2.4cm}|p{2.4cm}|}
% \hline
% \textbf{Attack Class} & \textbf{NSL-KDD Total Instances} & \textbf{KDD Cup '99 Total Instances} \\ \hline
% \textbf{DoS}          & 53,387                          & 3,883,370                            \\ \hline
% \textbf{Probe}        & 14,077                          & 41,102                               \\ \hline
% \textbf{R2L}          & 4,833                           & 1,126                                \\ \hline
% \textbf{U2R}          & 119                             & 52                                   \\ \hline
% \textbf{Normal}       & --                              & 492,708                              \\ \hline
% \textbf{Total}        & 72,416                          & 4,418,358                            \\ \hline
% \end{tabular}
% %\hspace{1cm}
% \label{tab:combined-dataset-summary}
% \end{table}

\section{Methodology}

\subsection{Problem Definition}
We consider the development of a network attack detection system for an IoV system. In this system, we collect network data in the form of network flows, which are records of network key performance indicators (KPIs) aggregated over a period of time. In this scenario, we do not have prior knowledge or data collected of potential attacks, but do have network flows of benign traffic for a particular system. This presents an \textit{unsupervised} and \textit{unseen} attack detection problem. Given this challenge, we focus on the development of an attack detection system that is capable of learning the benign behavior with a high degree of accuracy and detecting deviations from that benign behavior.

A multivariate time series is a sequence of vectors observed at successive time points. Consider $\textbf{X} = (x_1, x_2, x_3, ..., x_N) \in \mathbb{R}^{m \times n}$, where $X$ is a collection of univariate time series, $m$ is the length of the multivariate time series, and $n$ is the number of variables. $\textbf{X}$ contains a sequence of $m$ feature vectors, $x_t \in \mathbb{R}^{m}$. In an IoV network, we consider that collected network KPI flows are such sequential multivariate time series that may vary in duration, leading to unevenly spaced data. Here, while the flows themselves may be unevenly spaced in this way, we consider that we can take equal-sized sequential collections of the flows to perform benign behavior base-lining and anomaly detection. We do this division of data based on several considerations. Firstly, we consider the case such that IoV networks may be computational resource limited and unable to perform continuous real-time inference on every network flow collected. Secondly, given the typical sustained duration of attacks upon a system, it is unlikely a singular flow will be representative of an entire attack. If the majority of attacks within a flow can be considered as anomalous, the sequence should be flagged. Similarly, we consider that we can better learn overall network behavioral patterns when looking at flows collected over periods of time. This is particularly important when attempting to establish short-term and long-term benign traffic behaviors.

% This system model is visualized in [make this figure] 
% system model necesscary?
\subsection{Analytical Framework for Cybersecurity Threat Models}\label{sec:analyticalframework}
To effectively design and evaluate anomaly detection mechanisms, it is crucial to formally define the attack models the system aims to defend against. In this context, we provide mathematical formulations for the primary cyber threats considered in this study: Brute Force Attacks, Denial of Service (DoS) Attacks, and Reconnaissance Attacks. These formal definitions establish a rigorous foundation for analyzing the model's detection capabilities and guiding the development of robust defense strategies.\\
%The proposed Joint Reconstruction-Triplet Loss Autoencoder leverages both reconstruction loss and triplet loss to effectively detect these attack scenarios. The reconstruction loss is sensitive to abrupt deviations in data, making it suitable for detecting attacks that cause sudden changes, while the triplet loss enhances the model's ability to differentiate between subtle variations typical of reconnaissance behaviors.\\
A brute force attack systematically attempts all possible combinations to guess passwords or cryptographic keys. Let:
\begin{itemize}
    \item $N$ be the total number of possible combinations: $N = A^k$, where $A$ is the alphabet size and $k$ is the password length.
    \item $T$ be the time to attempt one guess.
    \item $p$ be the number of parallel processors used.
\end{itemize}
The expected time to success is:
\begin{equation}
    T_{\text{avg}} = \frac{N}{2p} \times T
\end{equation}
The probability of success after time $t$ is:
\begin{equation}
    P_{\text{success}}(t) = \frac{r \times t}{N}
\end{equation}
where $r = \frac{1}{T}$ is the guessing rate.\\
A DoS attack overwhelms system resources, making services unavailable. Let:
\begin{itemize}
    \item $C$ be the system's capacity.
    \item $R_{\text{legit}}$ be the legitimate request rate.
    \item $R_{\text{attack}}$ be the attack request rate.
\end{itemize}
The system overloads when:
\begin{equation}
    R_{\text{legit}} + R_{\text{attack}} > C
\end{equation}
The probability of overload in a queuing model is:
\begin{equation}
    P_{\text{overload}} = \frac{\lambda_{\text{legit}} + \lambda_{\text{attack}}}{\mu}
\end{equation}
where $\mu$ is the service rate.\\
Recon attacks gather system information for future exploitation. Let:
\begin{itemize}
    \item $N$ be the number of IPs, $P$ the number of ports, and $S$ the number of services.
    \item $r_{\text{scan}}$ be the scanning rate.
    \item $d$ be the detection threshold.
\end{itemize}
The total search space is:
\begin{equation}
    \Omega = N \times P \times S
\end{equation}
The probability of detection over time is:
\begin{equation}
    p_{\text{detect}}(T) = 1 - e^{-\beta \cdot r_{\text{scan}} \cdot T}
\end{equation}
The probability of successfully finding a vulnerability is:
\begin{equation}
    P_{\text{success}} = 1 - \left( \frac{V - v}{V} \right)^{r_{\text{scan}} \cdot T}
\end{equation}

% The integration of the reconstruction loss and triplet loss (see Section \ref{sec:autoencoder architecture}) within the proposed model effectively addresses the diverse nature of cyber threats in IoV networks. The \textit{reconstruction loss} is particularly effective in detecting abrupt deviations caused by high-intensity attacks such as Brute Force and DoS, where system behaviors shift drastically. In contrast, the \textit{triplet loss} improves the model's ability to capture and differentiate subtle deviations typical of Reconnaissance attacks, where gradual and stealthy scanning activities occur. This complementary design ensures comprehensive detection performance against both overt and covert attack strategies.

\subsection{Rationale for Joint Autoencoder Based on Analytical Framework}

The analytical framework presented in Section \ref{sec:analyticalframework} categorizes cyber threats into dimensions with distinct characteristics, such as high-intensity DoS attacks and stealthier reconnaissance activities. Given the significant impacts that these diverse attack types can cause, particularly when trying to protect new and emerging systems that do not possess significant labeled attack data, a motivated technical approach that can adequately address the attacks detailed in the analytical framework is needed.

Our methodology is based on two primary assumptions from our framework: 1) Different attacks will present as statistically distinguishable deviations from learned benign network patterns that have been captured in sequential network flows. 2) The robust detection of zero-day attacks or detection of attacks in new environments with no preexisting attack data calls for the use of unsupervised learning based on benign data. This means that any method used must be trained on benign data, focusing on deviations from benign data performance as opposed to specific attack signatures.

Based on these assumptions, we propose a joint reconstruction error-contrastive loss approach that is designed to detect the threats outlined in our analytical framework. A typical reconstruction loss ($\mathcal{L}_{REC}$) is used to train the autoencoder on benign data samples only to model and accurately reconstruct normal network patterns. Sequences with significant deviations from the benign network traffic patterns, such as the abrupt change in traffic patterns originating from DoS attacks, will result in a high reconstruction error of those samples because the model is only trained to reconstruct samples from the benign data distribution.

However, being able to adapt to shifts in benign data distributions over time so benign data points are not misclassified as anomalies, calls for the additional of elements that can provide fine-grain discrimination to these subtle changes. To achieve this, we utilize a contrastive loss (here, triplet margin loss, denoted as $\mathcal{L}_{TML}$). This loss explicltly encourages the model to learn more discriminative latent space representations by causing separation between benign representations that were collected at different points in time during a system's operation while clustering similar variations (versions of benign samples augmented with subtle noise) closer together. This helps diversify the latent space of the benign representations in a way that reconstruction loss alone might overlook, allowing for the highest reconstruction errors to be isolated on true attack samples. By combining these two losses, our model is designed to effectively monitor for a wide range of anomalous behaivors that address the different dimensions of our threat framework to promote robustness for unseen attack detection. The specific technical details for implementing this model are detailed in Section \ref{sec:autoencoder architecture}.

\subsection{Autoencoder Architecture}
\label{sec:autoencoder architecture}
\begin{figure*}[ht!]
    \centering
    \includegraphics[scale=0.37]{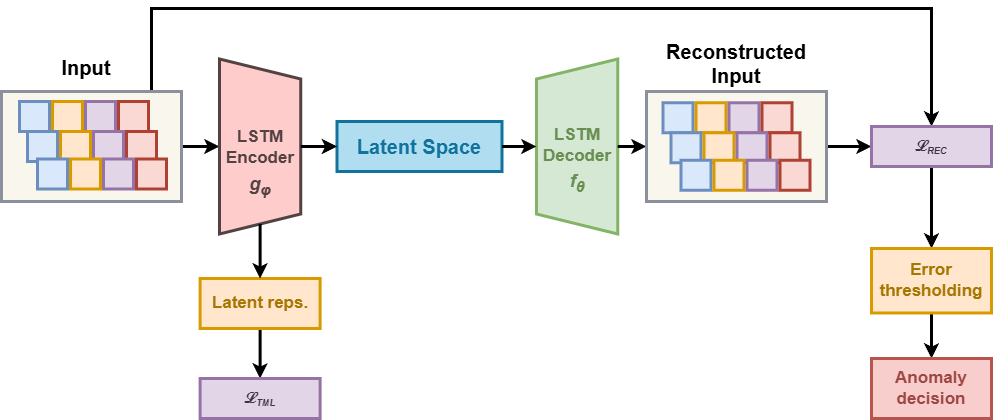}
    \makeatletter
    \renewcommand{\@makecaption}[2]{\centering \small #1. #2\par}
    \makeatother
    \caption{Proposed joint triplet-reconstruction loss autoencoder architecture utilizing LSTM layers.}
    \label{fig:modelfigure}
\end{figure*}

We opt for an autoencoder-based approach to the anomaly detection task. Specifically, we develop an autoencoder with a traditional encoder-decoder architecture designed to deal with a multivariate time-series data from collected IoT network flows. Our architecture focuses on the loss values and data structures necessary for the robust detection of network attacks. The architecture is depicted in Figure \ref{fig:modelfigure}.

\subsubsection{Encoder-decoder architecture} Here, we utilize LSTM layers for both the encoder, $g_{\phi}$ and decoder, $f_{\theta}$. We add an additional linear layer at the output of the decoder.

\cite{michelucci2022introduction} provides an introduction to the larger formulations behind autoencoders. We summarize the general representations of the encoder-decoder structure here. We can generically represent a encoder by the formulation:
\begin{equation}
    h_i = g_{\phi}(x_i) 
\end{equation}
where $h_i$ is the latent feature representation of sample $x_i$ generated by the encoder $g_{\phi}$. From this, we can generally represent a decoder by:
\begin{equation}
    \hat{x_i} = f_{\theta}(h_i) = f_\theta(g_\phi(x_i))
\end{equation}
where ${\hat{x_i}}$ is the reconstructed input generated by the decoder, $f_\theta$. 

\subsubsection{Losses} Our interest strongly lies in the loss functions utilized for the autoencoder-based approach for anomaly detection. Given that our application domain is one that is entirely unsupervised, we are unable to directly use some supervised training methodologies or loss functions. This assumption of a limited scope of data guides our training process as we aim to design a method that can accurately learn the benign representations without over-fitting such that the model is too sensitive to small benign changes or too agnostic to changes indicating anomalous behavior.

Traditionally, autoencoders have trained on a reconstruction loss, $\mathcal{L}_{REC}$ such that the model is tuned to the reconstruction task. This loss can be defined generally as:
\begin{equation}
    \mathcal{L}_{REC} = \frac{1}{N} \sum_{i=1}^{N} \mathcal{L}(x_i,\hat{x_i} )
\end{equation}
where $x_i$ is a input sample, $\hat{x_i}$ is the reconstructed sample, $\mathcal{L}$ is the chosen loss function for reconstruction loss, and $N$ is the number of samples. The chosen loss function can vary based upon task, but is often the mean squared error (MSE) loss or mean absolute error (MAE) loss.

While we still utilize the reconstruction task and loss for anomaly detection, we consider that the benign data may have temporal fluctuations that lead to false-positive results for benign network behavior sequences. In an attempt to diversify the latent space while taking into account the necessity for clear boundaries between benign and anomalous behavior, we include a triplet margin loss factor, $\mathcal{L}_{TML}$, in our training process to help guide the training process. This loss has been used in the vision domain for tasks such as face recognition \cite{xieface,WANGface,Schroff_2015_CVPR}, but we adopt it for the time-series classification task in this work. It is defined as:
\begin{equation}
    \mathcal{L}_{TML} = max\{d(a_i,p_i) - d(a_i,n_i) + m, 0\}
\end{equation}
where $m$ is the margin value, $a_i$ is an anchor sequence, $p_i$ is a positive example for the anchor sequence, and $n_i$ is a negative example for the anchor sequence. $d(x_i,y_i)$ is defined as:
\begin{equation}
    d(x_i,y_i) = \|x_i - y_i\|_2
\end{equation}

Here, we specifically feed the latent representations of the anchor, positive, and negative sequences produced from the encoder to our triplet margin loss in order to isolate on the purpose of latent space diversification.

In this unsupervised setting, we do not want to utilize samples from the anomaly class for the negative samples in the triplet margin loss. Because we are utilizing the triplet margin loss to help the model capture intra-class variability, we opt for the positive samples to be versions of the anchor sample augmented with random noise, $\epsilon$. This helps to learn the diversity of samples within the class while ensuring that the resulting positive sequence is somewhat similar to the anchor. For our negative samples, we select a different sequence within the benign set. We consider the value of this selection for the negative to be that while the negative is in the same class, it is a distinct different sequence and forces the model to learn the more fine-grained distinctions within the benign set. Through our sequencing methodology, we also ensure that there is a temporal difference between the anchor and negative sequence by picking another sequence in the set. We can leverage the similarities of samples for the positive samples generated while attempting to find the fine-grained distinctions between those similarities with the negative class.

We also experiment with weighting the loss terms using weights denoted as $\lambda_{REC}$ and $\lambda_{TML}$. Our overall loss term is then defined as:
\begin{equation}
    \mathcal{L} = \lambda_{TML}*\mathcal{L}_{TML} + \lambda_{REC}*\mathcal{L}_{REC}
\end{equation}

\subsection{Anomaly detection thresholding} In line with previous autoencoder-based anomaly detection methods, we train the autoencoder on benign data samples. After training, we first set how an anomaly is detected. Based on the training set, a reconstruction error threshold is set. We note here that we only utilize the reconstruction error in the threshold selection and we do not scale it with the $\lambda_{REC}$ value.

As opposed to setting the threshold to a static number, we utilize percentile values of the reconstruction errors generated from the benign set to allow for flexibility in threshold value based on training performance. We provide analysis results for percentile values between 90\% and 100\% to illustrate how percentile values may impact performance and set the percentile value to 99\% in all other trials.

If the reconstruction error of a sample is below this threshold, the sample is benign. Otherwise, the sample is considered anomalous. Here, we utilize the L2 norm/mean squared error (MSE) as our reconstruction error metric.

\section{Evaluation}

\subsubsection{Pre-processing} 
\begin{figure}[ht!]
    \centering
    \includegraphics[width=0.85\linewidth]{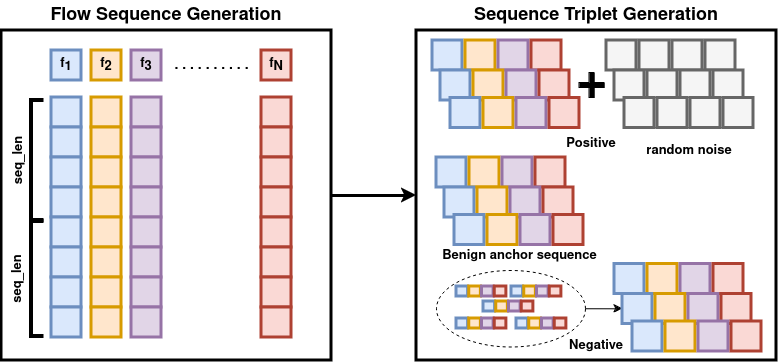}
        \makeatletter
    \renewcommand{\@makecaption}[2]{\centering \small #1. #2\par}
    \makeatother
    \captionsetup{justification=centering} 
    \caption{Sequencing for IoT network flows}
    \label{fig:sequencing}
\end{figure}

We utilize Min-Max normalization on both datasets. Given the data imbalance favoring the percentage of the ACI attack data, which is an inverse of the traditional imbalance problem within anomaly detection tasks, we investigate the inclusion and exclusion of the SMOTE \cite{smote} oversampling technique. When SMOTE is utilized, we are specifically oversampling the original network flows prior to sequence building. Outside of the sequence length variation trials, we fix the sequence length to 25 for all tables. For triplet generation, we fix the noise augmentation for positive sequence generation to 0.01. The triplet building process is visualized in Figure \ref{fig:sequencing}.

\subsubsection{Implementation Details}
For both datasets, we utilize 80\% of the benign set for training purposes and 20\% of the benign set for testing benign data. We sweep across the range of $\lambda_{TML} = [0, 1]$ and $\lambda_{REC} = [0, 1]$ in increments of 0.1 to identify the ideal combinations of these weights and present the results for these ideal values unless otherwise noted.

\subsection{Accuracy metricing}
To provide a comprehensive understanding of the model's performance, we formally define the core evaluation metrics used in this study. These metrics are essential to accurately assess the ability of the model to distinguish between benign and anomalous behaviors.\\
\textbf{Benign Accuracy (BA):} Measures the model's ability to correctly classify benign (normal) traffic.
\begin{equation}
    \text{BA} = \frac{\text{True Negatives (TN)}}{\text{True Negatives (TN)} + \text{False Positives (FP)}}
\end{equation}
\textbf{Anomaly Accuracy (AA):} Measures the model's ability to correctly detect anomalous (attack) traffic.
\begin{equation}
    \text{AA} = \frac{\text{True Positives (TP)}}{\text{True Positives (TP)} + \text{False Negatives (FN)}}
\end{equation}
\textbf{Precision (P):} Represents the proportion of correctly identified anomalies among all detected anomalies.
\begin{equation}
    \text{Precision} = \frac{\text{True Positives (TP)}}{\text{True Positives (TP)} + \text{False Positives (FP)}}
    \label{eq:p}
\end{equation}
\textbf{Recall (R):} Represents the proportion of detected anomalies among all actual anomalies.
\begin{equation}
    \text{Recall} = \frac{\text{True Positives (TP)}}{\text{True Positives (TP)} + \text{False Negatives (FN)}}
    \label{eq:r}
\end{equation}
\textbf{F1 Score (F1):} Provides a balance between Precision and Recall by computing their harmonic mean.
\begin{equation}
    \text{F1} = 2 \times \frac{\text{Precision} \times \text{Recall}}{\text{Precision} + \text{Recall}}
    \label{eq:f1}
\end{equation}

We provide brief benchmarking results on an unsupervised ML method to illustrate why ML is insufficient for application scenario of having no attack data in the training set. We leverage the isolation forest ML algorithm, which is designed to isolate anomalies from the decision trees it generates for each dataset. Table \ref{tab:datasets_onlybenign} gives the results for this scenario. We observe, in consideration of the benign-anomaly imbalances, near-random accuracy performance and imbalanced precision-recall values for both datasets. This performance motivates the need to develop more sophisticated models, as proposed in this paper.

\begin{table}[ht!]
    \centering
    \caption{Isolation Forest results, trained only on benign samples}
    \begin{tabular}{|c|c|c|c|}
        \hline
         \textbf{Dataset} &  \textbf{Accuracy} & \textbf{Precision} & \textbf{Recall}\\ 
         \hline
         \textbf{ACI-2023} & 13.3338 & 0.8984 & 0.0952\\
         \hline
         \textbf{WUSTL-2021} & 91.6900 & 0.4662 & 1.0000\\
         \hline
    \end{tabular}
    \label{tab:datasets_onlybenign}
\end{table}

Additionally, we utilize PyOD \cite{zhao2019pyod}, a Python toolbox used for detecting anomalies in multivariate data, for the intrusion detection task on our utilized datasets. PyOD provides a variety of state-of-the-art outlier detection models for benchmarking outlier detection model performance. We provide results for a Deep One-Class Classifier with AutoEncoder (DeepSVDD) \cite{deepsvdd} model and a Gaussian Mixture Model (GMM) \cite{GMM_source} in Table \ref{tab:performanceacrossall}. These models are trained in the same manner as our approach, where we use only benign samples for training and both benign and attack samples for testing.

We also provide results on an autoencoder utilizing only reconstruction error during the training process to highlight why our modifications to the traditional autoencoder are necessary. Table \ref{tab:performanceacrossall} presents the results for both the WUSTL and ACI datasets. While we found that a traditional approach could perform well on the WUSTL dataset, the model was unable to correctly capture the ACI dataset's benign behavior leading to poor anomaly detection accuracy results. This highlighted that the reconstruction error approach alone was not adaptable to other IoV representative datasets, which was a key consideration for our work. Additionally, we found that our proposed method could help boost the WUSTL dataset's accuracy values, as illustrated in the following sections. 

% \begin{table}[ht!]
%     \centering
%     \caption{\blue{Traditional reconstruction error-only autoencoder approach results}}
%     \begin{tabular}{|c|c|c|c|c|}
%         \hline
%          \textbf{Dataset} &  \textbf{Benign Acc.} & \textbf{Anom Acc.} & \textbf{Precision} & \textbf{Recall}\\ 
%          \hline
%          \textbf{ACI-2023} & 99.94 & 66.64 & 0.6664 & 0.9999\\
%          \hline
%          \textbf{WUSTL-2021} & 94.97 & 99.99 & 0.9999 & 0.9975\\
%          \hline
%     \end{tabular}
%     \label{tab:datasets_onlybenign_traditionalae}
% \end{table}

\begin{table*}[ht!]
    \centering
    \caption{Accuracy metrics for ACI-2023 and WUSTL-2021 datasets across different models.}
    \begin{tabular}{|c|c|c|c|c|c|c|c|c|c|c|}
        \hline
        & \multicolumn{5}{|c|}{\textbf{ACI-2023}} & \multicolumn{5}{|c|}{\textbf{WUSTL-2021}} \\
        \hline
        & \textbf{AE} & \textbf{Joint VAE} & \textbf{Joint AE} & \textbf{DeepSVDD \cite{deepsvdd}} & \textbf{GMM \cite{GMM_source}} & \textbf{AE} & \textbf{Joint VAE} & \textbf{Joint AE} & \textbf{DeepSVDD \cite{deepsvdd}} & \textbf{GMM \cite{GMM_source}}\\
        \hline
        \textbf{Benign Acc.} & \textbf{99.94} & 96.22 & 99.06 & 88.31 & 88.31 & 94.97 & \textbf{99.13} & 99.09 & 99.96 & 100.00\\
        \hline
        \textbf{Anomaly Acc.} & 66.64 & 96.91 & \textbf{97.28} & 77.80 & 76.82 & 99.99 & \textbf{100.00} & \textbf{100.00} & 90.07 & 90.10\\
        \hline
        \textbf{Precision} & 0.6664 & 0.9691 & \textbf{0.9729} & 0.9994 & 0.9994 & 0.9999 & \textbf{1.0000} & 0.9775 & 0.7982 & 0.7988 \\
        \hline
        \textbf{Recall} & 0.9999 & 0.9998 & \textbf{1.0000} & 0.7780 & 0.7682 & \textbf{0.9975} & 0.9784 & 0.9886 & 0.9960 & 1.0000\\
        \hline
        \textbf{F-one} & 0.7998 & 0.9843 & \textbf{0.9862} & 0.8749 & 0.8687 & \textbf{0.9987} & 0.9891 & 0.9830 & 0.8876 & 0.8881\\
        \hline
    \end{tabular}
    \label{tab:performanceacrossall}
\end{table*}

\subsubsection{ACI} Table \ref{tab:performanceacrossall} shows the benign test set accuracy, anomaly set accuracy, overall precision, and overall recall values for the ACI dataset given the identified ideal $\lambda_{TML}, \lambda_{REC}$ pair. We find that our joint AE approach outperforms all baselines in every metric except for benign accuracy, but we still see 99\% accuracy in this metric.

% \begin{table}[ht!]
%     \centering
%     \caption{\blue{ACI (no SMOTE) metrics for joint autoencoder. $\lambda_{REC} = 0.7, \lambda_{TML} = 0.9$, threshold = 99th percentile}}
%     \begin{tabular}{|c|c|c|c|}
%         \hline
%          \textbf{Benign Acc. }& \textbf{Anom. Acc.} & \textbf{Precision} & \textbf{Recall}\\
%          \hline
%          99.0566 & 97.2868 & 0.9729 & 1.0000\\
%          \hline
%     \end{tabular}
%     \label{tab:aci_nosmote_seq_model_best_weight_scores}
% \end{table}

Table \ref{tab:aci_smote_or_no_smoke_scores} breaks down the classification accuracy for ACI, without and with SMOTE, across attack categories. That is, given the three overall attack days/categories of brute force, DoS, and reconnaissance, we evaluate the individual accuracy for each category. We see the highest performance for the brute force attacks at 100\% detection accuracy and the worst performance for the DoS attacks at 91\% detection accuracy. We note that we have the same benign accuracy values from Table \ref{tab:performanceacrossall}. Given the lower performance for DoS-specific detection, we also explore if utilizing SMOTE in our training process increases the multi-category detection performance of our model. Table \ref{tab:aci_smote_or_no_smoke_scores} shows the accuracy metricing for the multi-category detection task utilizing SMOTE. We note that while there are some minimal decreases in the metrics for the brute force and reconnaissance attacks, there is a 4\% increase in the overall accuracy for DoS attack detection and a subsequent increase in the precision value for the DoS attacks.

% \begin{table}[ht!]
%     \centering
%     \caption{ACI (no SMOTE) metrics for multi-class classification, by overall attack type. $\lambda_{REC} = 0.6, \lambda_{TML} = 1.0$, threshold = 99th percentile}
%     \begin{tabular}{|c|c|c|c|c|}
%         \hline
%          \textbf{{Category}} & \textbf{Anom. Acc.} & \textbf{Precision} & \textbf{Recall}\\
%          \hline
%          Brute Force & 100.0000 & 1.000 & 0.9961\\
%          \hline
%          DoS & 91.7481 & 0.9175 & 0.9998 \\
%          \hline
%          Recon & 99.0946 & 0.9909 & 1.0000\\
%         \hline
%     \end{tabular}
%     \label{tab:aci_nosmote_mult_scores}
% \end{table}

\begin{table}[ht!]
    \centering
    \caption{ACI metrics for multi-class classification, with and without SMOTE.}
    \begin{tabular}{|c|c|c|c|c|}
        \hline
         \textbf{Category} & \textbf{SMOTE?} & \textbf{Anom. Acc.} & \textbf{Precision} & \textbf{Recall} \\
        \hline
         Brute Force & N & 100.0000 & 1.000 & 0.9961 \\
         \hline
         Brute Force & Y & 98.3408 & 0.9834 & 0.9915 \\
        \hline
         DoS & N & 91.7481 & 0.9175 & 0.9998 \\
        \hline
         DoS & Y & 98.3408 & 0.9834 & 0.9915 \\
        \hline
         Recon & N & 99.0946 & 0.9909 & 1.0000 \\
         \hline
         Recon & Y & 98.3408 & 0.9834 & 0.9915 \\
        \hline
    \end{tabular}
    \label{tab:aci_smote_or_no_smoke_scores}
\end{table}

\subsubsection{WUSTL} Table \ref{tab:performanceacrossall} shows the same metrics for the WUSTL dataset. Notably, the anomaly accuracy here is 100\%, but we reiterate that the number of anomaly samples is small within the WUSTL dataset compared to the benign set, which can lead to such high performance due to the imbalance.  Nonetheless, we see high performance on the precision and accuracy values as well for traffic within the WUSTL dataset. Here, we find that the AE-based models outperform the traditional multivariate detection schemes utilized from PyOD.

% \begin{table}[ht!]
%     \centering
%     \caption{\blue{WUSTL metrics for joint autoencoder. $\lambda_{REC} = 0.6, \lambda_{TML} = 1.0$, threshold = 99th percentile}}
%     \begin{tabular}{|c|c|c|c|}
%         \hline
%          \textbf{Benign Acc.}& \textbf{Anom. Acc.} & \textbf{Precision} & \textbf{Recall}\\
%          \hline
%          99.0970 & 100.0000 & 0.97752 & 0.9886\\
%          \hline
%     \end{tabular}
%     \label{tab:washuseq_model_best_weight_scores}
% \end{table}

\subsubsection{Impact of oversampling} Considering the imbalance of the ACI dataset representing an inverse imbalance compared to traditional anomaly detection datasets, it allows us to explore the role of oversampling in IoT network anomaly detection. Given an environment where benign samples may be limited for various reasons (adverse conditions, limited computation resources, etc), oversampling available data or generating synthetic samples can help diversify the benign training set. Here, we evaluate the use of Synthetic Minority Oversampling Technique (SMOTE) \cite{smote} to develop a more robust data distribution for the benign training data with our method. 

Table \ref{tab:aci_smote_seq_model_best_weight_scores} shows the overall accuracy metricing on ACI with and without utilizing SMOTE. We observe that resolving the benign imbalance issue with SMOTE helps boost the benign test, anomaly test, and overall precision values at the cost of a small decrease in overall recall.

\begin{table}[ht!]
    \centering
    \caption{ACI (SMOTE) metrics for joint autoencoder. $\lambda_{REC} = 0.8, \lambda_{TML} = 0.9$, threshold = 99th percentile}
    \begin{tabular}{|c|c|c|c|}
        \hline
         \textbf{Benign Acc. }& \textbf{Anom. Acc. } & \textbf{Precision} & \textbf{Recall}\\
         \hline
         99.9110 & 98.3408 & 0.9834 & 0.9915\\
         \hline
    \end{tabular}
    \label{tab:aci_smote_seq_model_best_weight_scores}
\end{table}

We also observe that utilizing oversampling helps overcome sensitivity to $\lambda_{REC}$ and $\lambda_{TML}$ values. In Table \ref{tab:aciaverageacross}, we see that the average accuracy metric values across all loss weight values but recall are higher when utilizing SMOTE.

\begin{table}[ht!]
    \centering
      \caption{ACI metrics average across all $\lambda_{REC}, \lambda_{TML}$, pairs (no SMOTE versus SMOTE), threshold = 99th percentile}
    \begin{tabular}{|c|c|c|c|c|}
        \hline
         \textbf{SMOTE?} & \textbf{Benign Acc. }& \textbf{Anom. Acc.} & \textbf{Precision} & \textbf{Recall}\\
         \hline
         Y & 99.7066 & 98.2166 & 0.9822 & 0.9994\\
         \hline
         N & 94.1081 & 95.7319 & 0.9573 & 0.9998\\
        \hline
    \end{tabular}
    \label{tab:aciaverageacross}
\end{table}

% \begin{table}[ht!]
%     \centering
%     \caption{\blue{ACI (SMOTE) metrics for multi-class classification, by overall attack type. $\lambda_{REC} = 0.8, \lambda_{TML} = 0.9$, threshold = 99th percentile}}
%     \begin{tabular}{|c|c|c|c|c|}
%         \hline
%          \textbf{Category} & \textbf{Anom. Acc.} & \textbf{Precision} & \textbf{Recall}\\
%          \hline
%          Brute Force & 100.0000 & 1.000 & 0.9884\\
%          \hline
%          DoS & 95.8103 & 0.9581 & 0.9995\\
%          \hline
%          Recon & 99.0946 & 0.9909 & 1.0000\\
%         \hline
%     \end{tabular}
%     \label{tab:aci_smote_mult_scores}
% \end{table}

We find that, overall, SMOTE is useful in developing a more robust benign training set when an IoV system's pre-collected benign data may be limited. In this specific application, given that there is a variety of environments and system configurations that may be present in IoV, we emphasize techniques such as SMOTE to help ensure the best system performance. This is also with a minimal amount of fine-tuning needed, represented by its reduction in sensitivity to the loss weights.

\subsubsection{Sequence Length Variations}

Given that our approach relies on the sequencing of data, it is relevant to discuss how variations in sequencing can impact the overall results and what the sequence length value represents in a practical setting. Table \ref{tab:aciaverageacross_seqlen} shows the results for sequence lengths of 10, 25, 50, and 100 on the ACI dataset. For fairness, we average across all loss weight pairs to present these results.

\begin{table}[ht!]
    \centering
      \caption{ACI metrics average across all $\lambda_{REC}, \lambda_{TML}$, pairs (varying sequence length), threshold = 99th percentile}
    \begin{tabular}{|c|c|c|c|c|}
        \hline
         \textbf{Length}& \textbf{Benign Acc. }& \textbf{Anom. Acc.} & \textbf{Precision} & \textbf{Recall}\\
         \hline
         10 & 96.4495 & 85.5202 & 0.8552 & 0.9998\\
         \hline
         25 & 94.1081 & 95.7319 & 0.9573 & 0.9998\\
         \hline
         50 & 99.3654 & 97.2389 & 0.9724 & 0.9999\\
         \hline
         100 & 96.2238 & 98.1101 & 0.9811 & 0.9999\\
        \hline
    \end{tabular}
    \label{tab:aciaverageacross_seqlen}
\end{table}

We find that, generally, the model performs better as we increase the sequence length, although there is some benign accuracy degradation between the 50 and 100 sequence lengths. However, a shorter sequence length translates to a higher periodicity in the detection system, which could have broader implications for overall system security. For more safety and time-critical systems, such as an IoV system, a higher periodicity that the compute capabilities of the systemc can support is preferred. We also note that while we use averaging for fairness, we still see high performing individual trials with small sequence sizes if the proper loss weights are utilized. For a sequence length of 10, the trial with $\lambda_{REC} = 0.6$ and $\lambda_{TML} = 0.7$ achieved 96\% benign accuracy and 98\% anomaly accuracy on the ACI dataset. Thus, while higher sequence lengths are less sensitive to the loss weights, lower sequence lengths can be balanced to fit an environment's specific needs. 

\subsection{Ablation Studies}
\subsubsection{Transfer Learning}
\label{sec:transfer_learning}

Given the importance of generalizability as it relates to attack detection across diverse environments, we also highlight our model's performance for the transfer learning task. In the IoV security setting, we consider that we may need zero-day detection systems for a given environment that has no pre-collected attack data. As such, leveraging the performance our model trained on data from a different environment on a new environment's data is a useful consideration of the adaptability of our model.

Given the different application domains of the WUSTL and ACI datasets, we explore the viability of training on the WUSTL dataset, freezing model weights, and fine-tuning on the ACI dataset. For our pre-trained cases, we set the triplet margin loss weight to 0 and the reconstruction loss to 1. Through this, we evaluate whether the latent space learning process introduced through the triplet margin loss for the WUSTL dataset positively impacts the reconstruction error for anomaly detection task.

Table \ref{tab:transfer_learning_inputoutput} shows the experimental results for transfer learning freezing all but the input and output layer WUSTL-trained weights for the model with the ideal loss weights from the WUSTL pre-training ($\lambda_{REC} = 0.6, \lambda_{TML} = 1.0$). Table \ref{tab:transfer_learning_encoder} shows the experimental results for transfer learning freezing the encoder weights for the model.

\begin{table}[ht!]
    \centering
    \caption{ACI (no SMOTE) metrics for joint autoencoder, pretrained on WUSTL and freezing all but input and output layers. $\lambda_{REC} = 0.6, \lambda_{TML} = 1.0$, threshold = 99th percentile}
    \begin{tabular}{|c|c|c|c|c|}
        \hline
         \textbf{PT?} & \textbf{Benign Acc.}& \textbf{Anom. Acc.} & \textbf{Precision} & \textbf{Recall}\\
         \hline
         Y & 99.0566 & 96.1830 & 0.9618 & 1.0000\\
         \hline
         N & 90.5661 & 97.4403 & 0.9755 & 0.9996\\
        \hline
    \end{tabular}
    \label{tab:transfer_learning_inputoutput}
\end{table}

\begin{table}[ht!]
    \centering
     \caption{ACI (no SMOTE) metrics for joint autoencoder, pretrained on WUSTL and freezing the encoder only. $\lambda_{REC} = 0.6, \lambda_{TML} = 1.0$, threshold = 99th percentile}
    \begin{tabular}{|c|c|c|c|c|}
        \hline
         \textbf{PT?} & \textbf{Benign Acc.}& \textbf{Anom. Acc.} & \textbf{Precision} & \textbf{Recall}\\
         \hline
         Y & 99.0566 & 96.3717 & 0.9637 & 1.0000\\
         \hline
         N & 90.5661 & 97.4403 & 0.9755 & 0.9996\\
        \hline
    \end{tabular}
    \label{tab:transfer_learning_encoder}
\end{table}

In both, we see an increase in the overall detection accuracy for the benign sets as well as an increase in the recall values while we see a decrease in the anomaly accuracy and the precision value. This indicates the transfer learning was valuable for decreasing the number of false negatives at the trade-off of increased false positives. Freezing only the encoder yields slightly higher results between the two pre-trained cases, indicating performance benefits from letting the decoder train fully on a specific environment's reconstruction task. These results establish our model's generalizability capabilities and show that no attack data is needed in the source or target environment to detect any attack type in our method.

\subsubsection{Variational autoencoder (VAE)}

Variational Autoencoders (VAEs) are generative models that learn a structured latent space by imposing a probabilistic distribution (typically Gaussian) on the latent variables. VAEs introduce stochasticity through reparameterization. This makes VAEs particularly useful for tasks anomaly detection. However, VAEs can struggle with producing sharp reconstructions compared to AEs because of the trade-off between reconstruction accuracy and latent space regularization. Additionally, if the prior distribution is poorly chosen or the KL divergence term is too dominant, VAEs may produce overly blurred outputs or fail to capture fine details, whereas AEs, being purely deterministic, often achieve better reconstruction fidelity. The chosen distribution is particularly important in our context as our attack datasets are not guaranteened to be inherently Gaussian and assuming a fixed distribution could skew detection results. However, due to their potential benefits, we perform an ablation study on the modification of our model architecture to a VAE, so we utilize contrastive, reconstruction, and regularization loss. We present the results in Table \ref{tab:performanceacrossall}. We find that the joint VAE performs well but does not outperform the joint AE for the ACI-2023 dataset. Conversely, we find it performs similarly and, in some metrics, better than the joint AE for the WUSTL-2021 dataset.

\begin{figure}
    \centering
    \includegraphics[width=0.90\linewidth]{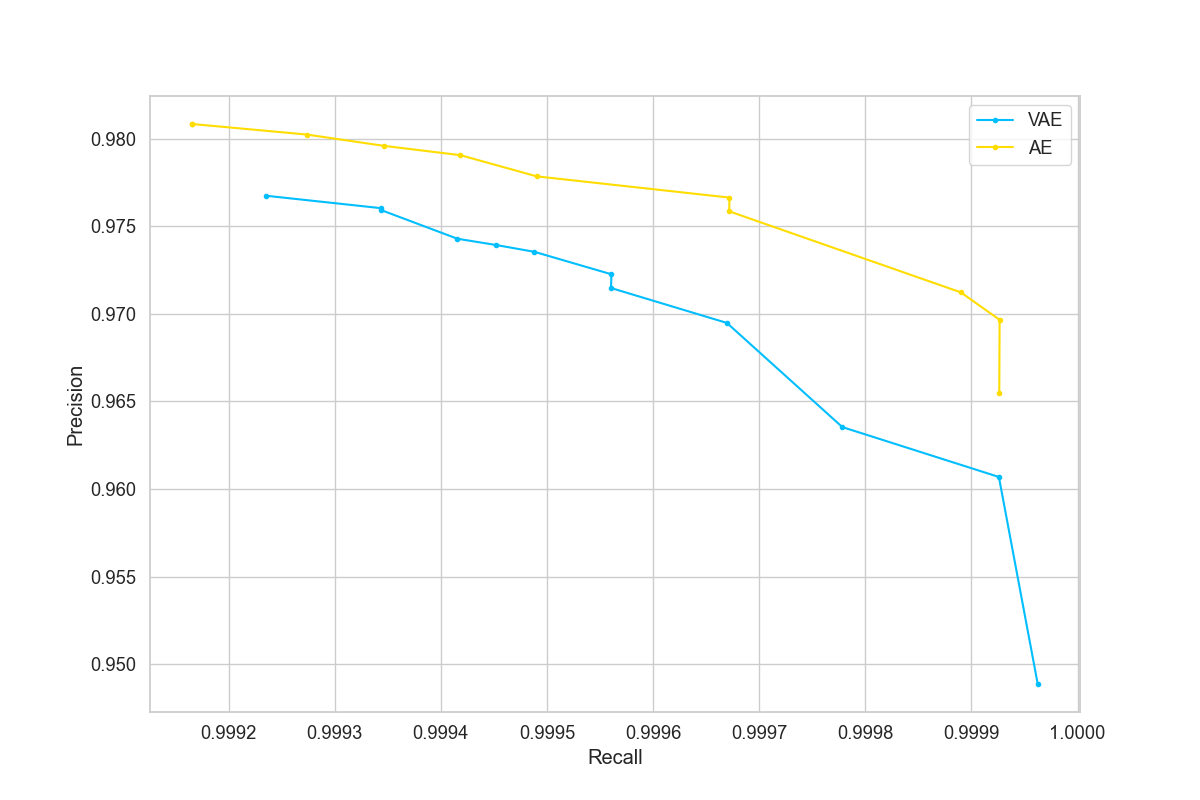}
    \makeatletter
    \renewcommand{\@makecaption}[2]{\centering \small #1. #2\par}
    \makeatother
    \caption{ACI precision-recall curve across percentile values for joint AE and joint VAE}
        \label{fig:aciprcurve}
\end{figure}

\subsubsection{Robustness Analysis Across Percentiles}
\begin{figure}
    \centering
    \includegraphics[width=0.90\linewidth]{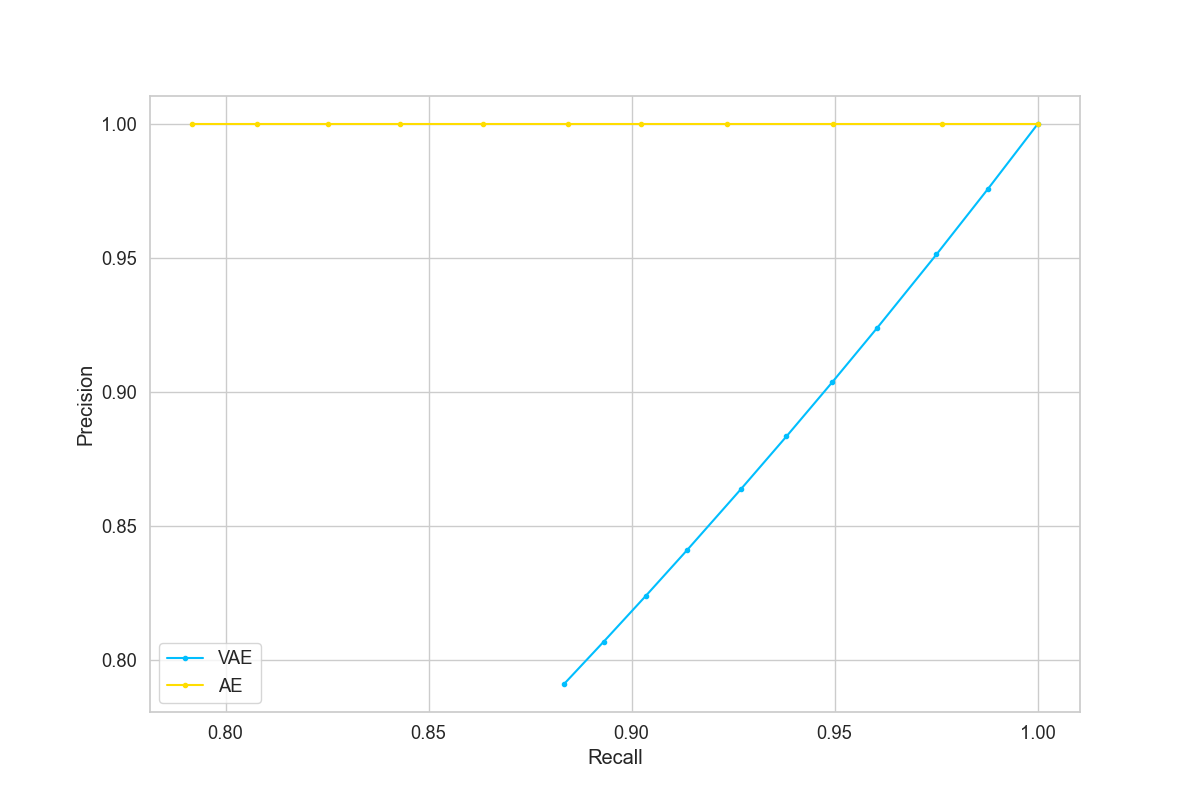}
    \makeatletter
    \renewcommand{\@makecaption}[2]{\centering \small #1. #2\par}
    \makeatother
    \caption{WUSTL precision-recall curve across percentile values for joint AE and joint VAE}
    \label{fig:wustlprcurve}
\end{figure}

To further evaluate model robustness, we analyze performance across different percentile thresholds of anomaly scores. By varying the decision thresholds, we assess how sensitive the model is to detecting anomalies at different operating points. This analysis provides insight into the model's stability and consistency, revealing how changes in the percentile threshold impact detection performance. We present these results numerically in Table \ref{tab:aci_differingpercentiles} for ACI and show corresponding precision-recall curves in Figure \ref{fig:aciprcurve} for ACI-2023 and Figure \ref{fig:wustlprcurve} for WUSTL-2021. We find that our overall benign accuracy increases dramatically as we increase our percentile while our other metrics have minor increases or decreases. Because the thresholding is intended to find the outliers from the reconstruction error distribution, as the percentile grows, only those extreme outliers should be flagged as anomalies. This leads to less benign samples being misclassified as attacks (ie, false positives).

% \begin{figure}
%     \centering
%     \includegraphics[width=\linewidth]{LaTeX/Figures/precision_recall_vae_ae_ACI.png}
%     \makeatletter
%     \renewcommand{\@makecaption}[2]{\centering \small #1. #2\par}
%     \makeatother
%     \caption{ACI precision-recall curve across percentile values for joint AE and joint VAE}
%         \label{fig:aciprcurve}
% \end{figure}

% \ts{Add Accuracy, Precision, and Recall for 90th,95th,99th percentile}

% \jb{will add this; to make precision-recall curves, I have all of them between 90 and 100 (100's redundant to use but still) so that can potentially be combined together}

\begin{table}[ht!]
    \centering
    \caption{ACI (no smote) metrics for joint autoencoder across differing percentile values}
    \begin{tabular}{|c|c|c|c|c|c|}
        \hline
         & \textbf{Benign Acc. }& \textbf{Anom. Acc. } & \textbf{Precision} & \textbf{Recall} & \textbf{F-one}\\
         \hline
         \textbf{90\%} & 78.30 & 98.08 & 0.9808 & 0.9992 & 0.9899\\
         \hline
         \textbf{95\%} & 86.79 & 97.79 & 0.9779 & 0.9994 & 0.9886\\
         \hline
         \textbf{99\% }& 99.06 & 97.29 & 0.9729 & 1.000 & 0.9862\\
         \hline
    \end{tabular}
    \label{tab:aci_differingpercentiles}
\end{table}

\begin{figure}
    \centering
    \includegraphics[width=0.85\linewidth]{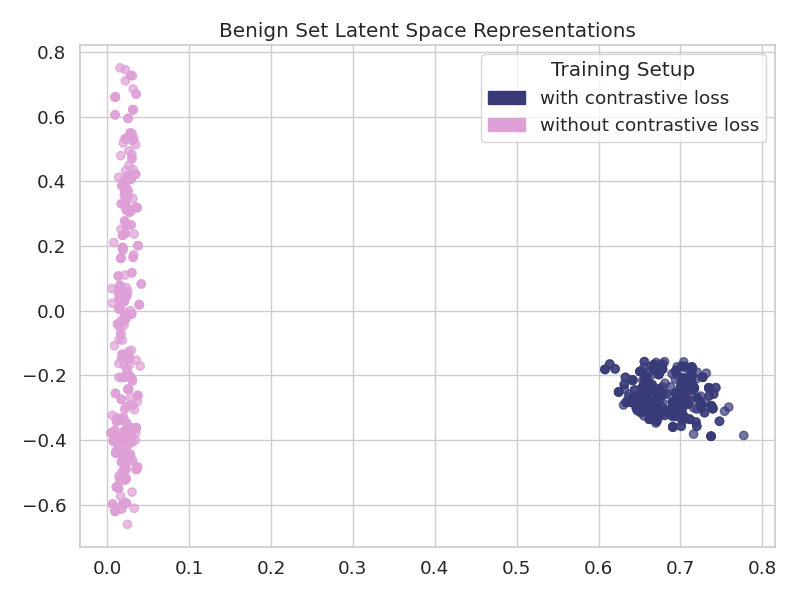}
     \makeatletter
    \renewcommand{\@makecaption}[2]{\centering \small #1. #2\par}
    \makeatother
    \caption{Benign representations with and without contrastive loss.}
    \label{fig:benign_fragmentation_con_loss}
\end{figure}

\subsubsection{Impact of Contrastive Loss on Benign Representations}Given our utilization of contrastive loss with the intent of creating better cohesion among benign representation samples, we explicitly highlight that the use of contrastive does help to boost benign representation cohesion. To evaluate this, we train our joint autoencoder method with and without contrastive loss. For fairness, we train the model with the same $\mathcal{L}_{REC}$ value (0.9) for both models. We then plot the benign representations derived from these two models in the latent space to observe with what level of cohesiveness the individual clusters of points are grouped. This is plotted in Figure \ref{fig:benign_fragmentation_con_loss}. Here, we qualitatively see that the latent space representations for benign data produced with contrastive loss have better cohesion around the center of that cluster of points. Quantitatively, we can use a measure of the average length along the axes to capture how wide the plotted points are in all directions. These results are provided in Table \ref{tab:avglength}. We find that the overall average length is smaller for the benign representations created using contrastive loss, indicating better cohesion.

\begin{table}
    \centering
        \caption{Average length along axes for benign representations}
    \begin{tabular}{|c|c|}
        \hline
         \textbf{With contrastive} loss & 0.2000\\
         \hline
         \textbf{Without contrastive loss} & 0.7244\\
         \hline
    \end{tabular}
    \label{tab:avglength}
\end{table}

\section{Conclusion}

Here, we presented a unique autoencoder method for network attack detection in IoV environments. We conducted extensive metricing on two state-of-the-art datasets that are well representative of modern networking patterns in distributed networked systems. We show that this method can achieve high benign and anomaly test accuracy while having no attack data within the training set. These results demonstrate our model's capabilities for unseen attack detection in IoV environments. Additionally, we show that our model works as an adaptable attack detection mechanism for detection across different environments, as proven by our transfer learning study. Our work has key implications for improving the security of these emerging and safety-critical automotive systems via the presentation of a highly robust unseen attack detection.

\bibliographystyle{IEEEtran}
\bibliography{ieeeiotj}

\end{document}